\documentstyle[talk102,epsfig]{article}

\def\Journal#1#2#3#4{{#1} {\bf #2}, #3 (#4)}

\def\NPB{{\em Nucl. Phys.} B}
\def\PLB{{\em Phys. Lett.}  B}

\def\ZPC{{\em Z. Phys.} C}

\def\beq{\begin{equation}}
\def\eeq{\end{equation}}
\def\bea{\begin{eqnarray}}
\def\eea{\end{eqnarray}}
\newcommand {\Gevcc}    {\mbox{${\rm {GeV}}/c^2$} }
\newcommand {\ee}         {\mbox{$\rm {e^{+}e^{-}}$} }
\newcommand {\ff}         {\mbox{$\mathrm{f}\overline{\mathrm{f}}$}}
\newcommand {\lept}       {\mbox{${\ell^{+}\ell^{-}}$} }
\newcommand {\mumu}       {\mbox{$\mu^{+}\mu^{-}$} }
\newcommand {\tautau}     {\mbox{$\tau^{+}\tau^{-}$} }
\newcommand {\Zzero}      {\mbox{$\rm{Z}$} }
\newcommand {\Gff}        {\Gamma_{\mathrm{f}\overline{\mathrm{f}}}}
\newcommand {\Gee}        {\Gamma_{\rm {ee}}}
\newcommand {\Ghad}       {\Gamma_{\mathrm{had}}}
\newcommand {\Ginv}       {\Gamma_{\mathrm{inv}}}
\newcommand {\Gll}        {\Gamma_{\ell\ell}}
\newcommand {\GF}      {G_{\mathrm{F}}}
\newcommand {\GZ}      {\Gamma_{\mathrm{Z}}}
\newcommand {\MZ}      {M_{\mathrm{Z}}}
\newcommand {\alfas}   {\alpha_s}
\newcommand {\alfmz}   {\alfas(\MZ^2)}
\newcommand {\MH}      {M_{\mathrm{H }}}
\newcommand {\Mt}      {M_{\mathrm{t}}}
\newcommand {\spol}    {\sigma^o_{\rm{had}}}
\newcommand {\Ree}      {R_{\mathrm{e}}}
\newcommand {\Rmu}      {R_{\mu}}
\newcommand {\Rtau}      {R_{\tau}}
\newcommand {\Rl}      {R_{\ell}}

\newcommand {\gahatf}      {g_{{A}_{\rm f}}}
\newcommand {\gvhatf}      {g_{{V}_{\rm f}}}
\newcommand {\Afb}     {A_{\mathrm{FB}}}
\newcommand {\Afbze}     {\rm{A}\!^{0,\,{\rm e}}_{\rm {FB}}}
\newcommand {\Afbzf}     {\rm{A}\!^{0,\,{\rm f}}_{\rm {FB}}}
\newcommand {\Afbzm}     {\rm{A}\!^{0,\,\mu}_{\rm {FB}}}
\newcommand {\Afbzt}     {\rm{A}\!^{0,\,\tau}_{\rm {FB}}}
\newcommand {\Afbzl}     {\rm{A}\!^{0,\,\ell}_{\rm {FB}}}

\newcommand{\cs}{\mbox{$\mathrm{cos}\,\theta$} }
\newcommand {\css}      {\mbox{$\cos^2\theta^{*}$} }
\newcommand {\cAe} {\mbox{$\cal A_{\rm e}$}}
\newcommand {\cAf} {\mbox{$\cal A_{\rm f}$}}
\newcommand {\roots}      {\sqrt{s}}

\begin{document}

\title{Z LINESHAPE AND FORWARD-BACKWARD ASYMMETRIES} 

\author{B. BLOCH-DEVAUX }

\address{CEA,DAPNIA/Service de Physique des Particules, CE-Saclay, 
F-91191Gif-sur-Yvette Cedex,FRANCE \\E-mail: Brigitte.Bloch@cern.ch}


\twocolumn[\maketitle\abstracts{ Preliminary but close to final results
on the Z Lineshape and Forward-Backward asymmetries from the four LEP
experiments are presented. Combined values extracted from ALEPH,DELPHI,L3 
and OPAL data recorded at energies around the Z pole are discussed.}]

\section{Introduction}

 Preliminary results of the four LEP experiments~\cite{fourl}, 
including the whole data set recorded around the Z pole from 1990 to 1995, 
are presented here.  
 Emphasis will be given to changes since the last report presented at Winter 
Conferences and experimental systematic errors evaluation. The discussion of 
the results as a test of the Standard Model can be found in another talk of 
this session~\cite{mg}.

\section{Lineshape parameters} 
 The parametrisation of the dominant contribution to the cross section at
energies around the Z resonance is based on a Breit-Wigner shape with
an s-dependent width. After deconvolution of pure QED corrections,
this is given in the ``improved Born approximation'' by
\begin{equation}
\label{Eq:linshape}
\sigma_{{\rm f}\bar{\rm f}}(s)=
      \sigma^0_{{\rm f}\bar{\rm f}}
      {s\GZ^2 \over \left(s-\MZ^2\right)^2+s^2\GZ^2/\MZ^2}\; ,
\end{equation}
where the parameters are the Z mass, $\MZ$, the width, $\GZ$, and
the pole cross section, $\sigma^0_{{\rm f}\bar{\rm f}}$.
The contributions from $\gamma$ exchange and from
$\gamma$--$\Zzero$ interference are small at
energies $\sqrt{s}=\MZ$ and are set to their Standard Model values.
                                                                     
The pole cross section can be written in terms of partial decay widths of the
initial and final states, $\Gee$ and $\Gff$,
\begin{equation}
\sigma^0_{{\rm f}\bar{\rm f}}
  = {12\pi\over\MZ^2}{\Gee\Gff\over\GZ^2}\,.
\end{equation}
The definition of the widths explicitly includes QED and QCD corrections,
\begin{equation}
\label{Eq:gam}
\Gff=\frac{\GF\MZ^3}{6\pi\sqrt{2}} N_c^{\rm f}
(\gvhatf^2+\gahatf^2)(1+\delta_{QED}+\delta_{QCD})~;
\end{equation}
$\gvhatf$ and $\gahatf$ are effective vector and axial vector couplings of
the $\Zzero$ to fermion species ${\rm f}$,
and $N_c^{\rm f}$, the QCD colour factor, is one for leptonic final states
and three for hadrons. Due to the effect of radiative corrections, these
effective couplings are complex numbers. 
                                                                               As the pole cross sections, $\sigma^0_{{\rm f}\bar{\rm f}}$,
have in common the statistical and systematic error from the luminosity
determination, it is more practical to use the ratios of pole
cross sections in the parametrisation of the leptonic channels,
\begin{equation}
\label{Eq:rl}
R_{\rm f}  =   \frac {\spol } {\sigma^0_{{\rm f}\bar{\rm f}} }
        \equiv  \frac{\Ghad} {\Gff} {\rm ~~~for~~ } \ff=\ee,\mumu,\tautau.
\end{equation}

The energy dependence of the leptonic Forward-Backward asymmetry
depends on the axial coupling which is mainly determined by the
leptonic width. Therefore,
the measurements of the leptonic Forward-Backward asymmetries
can be condensed into one single
parameter per lepton species in the final state, the pole asymmetry
$\Afbzf$. This is given by the following combinations of
effective couplings
\begin{equation}
\label{Eq:afb0}
\Afbzf   \equiv  {3\over 4} \cAe\cAf
\end{equation}
with:
\begin{equation}
\label{eqn-cAf}
\cAf\equiv\frac{2\gvhatf \gahatf}
{\gvhatf^{2}+\gahatf^{2}}\,.
\end{equation}
                     
The contributions from $\gamma$--$\Zzero$ interference,
imaginary parts of the photon vacuum polarisation, and the imaginary
parts of the effective couplings are set to their
Standard Model expectations and not included in the definition of $\Afbzf$.

\section{The fit procedure}
 Each experiment provides a set of nine parameters ($\MZ$, $\GZ$, $\spol$,
$\Ree$, $\Rmu$, $\Rtau$,$\Afbze$, $\Afbzm$, $\Afbzt$) fitted to
the whole set of cross sections and Forward-Backward lepton
asymmetries accumulated by the four LEP experiments during the LEP~1 period.
This is done with the latest versions of the computer codes
MIZA~\cite{MIZA} and ZFITTER~\cite{ZFITTER,WGPC}. 
These codes provide parametrisations of the fermion pair production
cross sections and Forward-Backward asymmetries at energies around
the $\Zzero$ resonance in terms of effective couplings and also
calculate pure QED corrections to full ${\cal O}(\alpha^2)$ with
exponentiation of the soft part
or with leading ${\cal O}(\alpha^3)$.

The measured cross sections and Forward-Backward asymmetries are
treated in a $\chi^2$ minimisation procedure to extract the nine parameters
and their errors and correlations. The full error matrix of the input
measurement is  constructed. The covariance matrix includes
the statistical and systematic errors as well as their correlations.

Under the assumption of lepton universality, the nine parameters set reduces to
a five parameters one ($\MZ$, $\GZ$, $\spol$, $\Rl$, $\Afbzl$).
The chosen parameters are almost uncorrelated and thus well adapted to fitting
and averaging procedures.Table~\ref{corr} gives a typical correlation matrix.
\begin{table}
\begin{center}
\caption{Typical correlation matrix for a five parameter fit.}\label{corr} 
\vspace{0.1cm}
\begin{tabular}{|c| c c c c c|} 
\hline 
       & $\MZ$ &  $\GZ$ & $\spol$ & $\Rl$ & $\Afbzl$ \\
\hline
 $\MZ$ &   1.  &   0.05 & -0.01   & -0.02 & 0.06     \\
 $\GZ$ &       &   1.   & -0.16   & 0.00    & 0.00       \\
 $\spol$ &     &        &  1.     & 0.14   & 0.00      \\
 $\Rl$ &       &        &         &  1.     &  0.01  \\
 $\Afbzl$ &    &        &         &         &  1.    \\
\hline
\end{tabular}
\end{center}
\end{table}

The only relevant correlations are between ($\GZ$ and  $\spol$) and ($\Rl$ and  $\spol$). They contain the sensitivity to the Standard Model parameters $\MH$ 
and $\alfas$.

\section{Statistics analysed}
 Analyses have been performed on data sets recorded from 1990 to 1995. Several
energy scans were archieved, first in 1990 and 1991 within $\pm 3 GeV$ of  
$\roots = \MZ$ value,then again in 1993 and 1995 (within $\pm 2 GeV$), while
1992 and 1994 were devoted to peak energy only.
In total the four LEP experiments accumulated 4 times 4 Millions of Z decays,
corresponding to an integrated Luminosity of $120~pb^{-1}$ at the peak energy
and $40~pb^{-1}$ off peak. The statistical error on $\spol$ is therefore 
about $0.05\%$ per experiment and about $0.15\%$  per experiment for leptonic
cross-sections.
As no more running at the Z peak energy is foreseen before the end of LEP era,
the results presented here contain the final word on statistical errors.
The details of the improvements achieved in the experimental systematic errors
will be discussed in the next sections.

\section{Experimental errors}
 Each experiment provides a set of nine (five) parameters values and the 
corresponding covariance matrix where common systematic error sources are
clearly identified. Their origine is of two types: first, uncertainties in the
theoretical calculations, dominated by the uncertainty on the small angle 
Bhabha cross section and the $t$-channel contribution to the wide angle 
Bhabhas, second, a common experimental error from the LEP beam energy 
uncertainties. All will be discussed below in the appropriate sections.
  
Most of the correlations are specified via detailed correlation matrices
constructed from the specific inter-dependencies of the various error
contributions. Particular care is given to the evaluation of the correlated
and uncorrelated contribution of each source between data sets of different 
years as well as data sets of the same year recorded at different center of 
mass energies.

\section{$\MZ$}

 The very precise relative error obtained for $\MZ$ (about $2.10^{-5}$) has
been achieved after the impressive work of the LEP Energy Calibration 
group\cite{LEPEC}. Starting with a 45~MeV error from a simple on-line Field 
Display in 1989, the breakthrough was obtained in 1991 with the use of the 
depolarization resonance method to calibrate the beam energy. Later this 
technique was applied at several energy points and other internal (Radio 
Frequency corrections) and external sources causing geometrical deformations 
of the LEP ring (terrestrial tides) were corrected for in 1992, leading to a 
7~MeV uncertainty. Today the ultimate precision on the beam energy achieved is 
1.5~MeV. A correlation matrix has been carefully built describing the 
contributions of the different terms between years and energy points of the 
same year. The mass measurement is directly affected by the uncertainty of the 
absolute energy scale, i.e., uncertainties correlated between energy points.
The combined result from the LEP experiments is given in Table~\ref{tab:tabmz}. 
%
%
\begin{table}
\begin{center}
\caption{$\MZ$ value per experiment and combined result in MeV.}
\label{tab:tabmz}
\vspace{0.1cm}
\begin{tabular}{|c |c|}
\hline
 ALEPH  &  $91188.4 \pm 3.1$ \\
DELPHI &   $91186.6 \pm 2.9$ \\
L3 &  $91188.3 \pm 2.9$ \\
OPAL & $91184.8 \pm 3.0$ \\
\hline
 LEP  & $91186.7 \pm 2.1$ \\
\hline
\end{tabular}
\end{center}
\end{table}                                                           
\section {$\GZ$}

\begin{figure}
\psfig{file=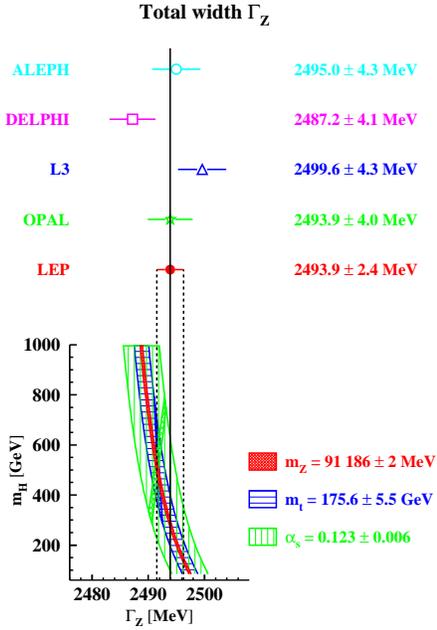,height=9.cm} 
\caption{$\GZ$ value for each LEP experiment and combined result.}
\label{fig:gz}
\end{figure}

 The Z width uncertainty is affected only by the error on the difference in 
energy between energy points (1.3 MeV), and not by the error on the absolute 
energy scale. An additionnal 0.2~MeV comes from the energy spread around 
the mean value. The combined result from the LEP experiments is given here 
and is shown in Figure~\ref{fig:gz}, the overall relative precision achieved 
is $0.1 \%$
\begin{equation}
\GZ = ( 2.4939 \pm 0.0019 \pm 0.015 ~(LEP)) \Gevcc.
\end{equation}

\section {$\spol$}

\begin{figure}
\psfig{file=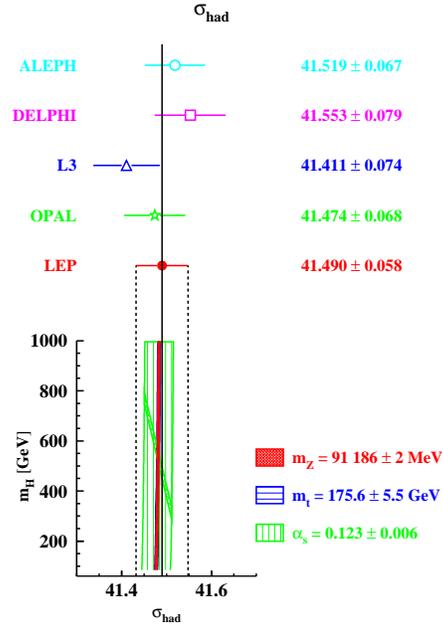,height=9.cm} 
\caption{Pole hadronic cross-section value for each LEP experiment and 
combined result.}
\label{fig:shad}
\end{figure}

 The main source of uncertainty comes here from the Luminosity error. All
four experiments have decreased their experimental error at the $0.1\%$ level.
The common theoretical error amounts to $0.11\%$. There are some hints of a 
further decrease in its evaluation~\cite{bward}. One should note that an
increased (with respect to reports at Winter Conferences) overall systematic 
error is quoted due to the theoretical 
uncertainty (0.021 nb) on the third order terms introduced in the lineshape 
description. The combined result from the LEP experiments is shown in 
Figure~\ref{fig:shad}.
\section {$\Rl$}

\begin{figure}
\psfig{file=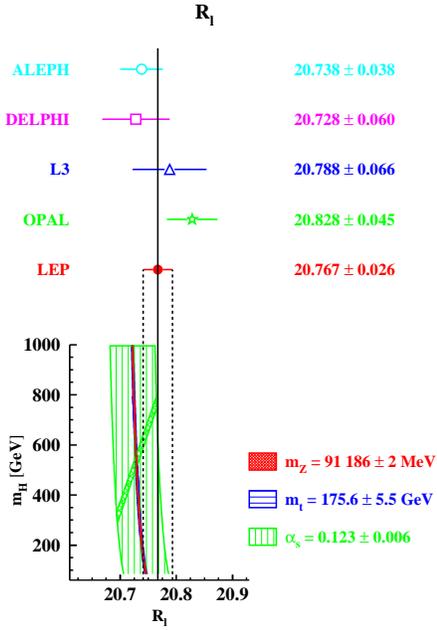,height=9.cm} 
\caption{$\Rl$ ratio value for each LEP experiment and combined result, 
assuming lepton universality.}
\label{fig:rl}
\end{figure}

 The errors on $\Ree$, $\Rmu$ and $\Rtau$ do not depend on luminosity but on 
the knowledge of relative acceptance and selection efficiencies. They are 
entirely dominated by the total error on the lepton cross sections, with very 
small correlations of about 5\% from the common normalisation to the number of
hadrons. Substantial improvements have been achieved by ALEPH and OPAL who
decreased the systematics uncertainty on $\Rl$ by almost a factor 2. Aleph has 
developped a new global lepton analysis which reduces the 
uncertainty on the flavour separation and is in very good agreement with the 
classical exclusive analysis. Therefore both analyses have been combined.
Particular care has been given to the $t$-channel subtraction, using the
prescription from the authors of ALIBABA~\cite{aliba}.  
The combined result from the LEP experiments assuming lepton universality is 
shown in Figure~\ref{fig:rl}.
The splitting into individual lepton species is given in Table~\ref{tab:rltab}

\begin{table}
\begin{center}
\caption{$\Rl$ ratio value for each experiment and each lepton specie.}
\label{tab:rltab}
\vspace{0.1cm}
\begin{tabular}{|c|c|c|c|c|}
\hline
  &  $\Ree$ &  $\Rmu$ &  $\Rtau$\\
\hline
 ALEPH  & $20.69 \pm 0.07$ &  $20.81 \pm 0.06$  &  $20.72 \pm 0.06$ \\
 DELPHI & $20.87 \pm 0.12$ &  $20.67 \pm 0.08$  &  $20.78 \pm 0.12$ \\
  L3    & $20.78 \pm 0.11$ &  $20.84 \pm 0.10$  &  $20.75 \pm 0.14$ \\
  OPAL  & $20.92 \pm 0.09$ &  $20.82 \pm 0.06$  &  $20.85 \pm 0.09$ \\
\hline
 LEP  & $20.78 \pm 0.05$ & $20.79 \pm 0.03$ & $20.76 \pm 0.05$\\ 
\hline
\end{tabular}
\end{center}
\end{table}

\section {$\Afbzl$}
\begin{figure}
\psfig{file=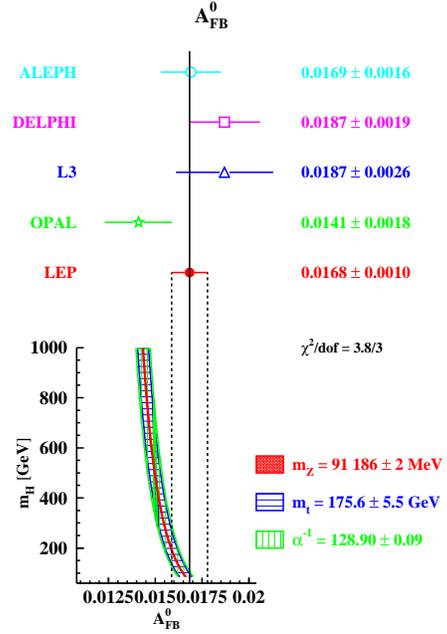,height=9.cm} 
\caption{$\Afbzl$ value assuming lepton universality, for each LEP experiment 
and combined result.}
\label{fig:Afbzl}
\end{figure}

 The differential cross section for the reaction $\ee\to\lept$ is expected to 
have an approximate quadratic dependence on the cosine of the polar angle:
\begin{equation}
\label{ds_dcost}
 \frac{d\sigma}{d\cs} \propto
\left( 1 + \css +  \frac{8}{3}\Afb \cs \right)\,,
\end{equation}
where $\theta$ is the angle between the incoming electron and the outgoing 
negative lepton.
The LEP experiments have finalized the lepton asymmetry analyses including 
periods in 1993,1994 and 1995 where precise LEP energy calibration was not
available but data could be used however for asymmetry measurements. 
In the case of muons and taus, the selection has been designed to minimize any 
final state invariant mass dependence as an acolinearity cut between the two 
final leptons is equivalent to an invariant mass  dependent  cut. For the 
electron final state, the final error is 
dominated by the uncertainty on the $t$-channel subtraction. The error 
correlation between the $\Rl$ and the $\Afbzl$ parameters is also taken into
account.
The combined result from the LEP experiments is shown in 
Figure~\ref{fig:Afbzl}, assuming lepton universality.
The splitting into individual lepton species is given in 
Table~\ref{tab:afbtab}.

\begin{table}
\begin{center}
\caption{$\Afbzl$ value for each experiment and each lepton specie.}
\label{tab:afbtab}
\vspace{0.1cm}
\begin{tabular}{|c|c|c|c|}
\hline
 Unit is $10^{-4}$   &  $\Afbze$ &  $\Afbzm$ &  $\Afbzt$\\
\hline
 ALEPH  & $181. \pm 33.$ &  $170. \pm 25.$  &  $166. \pm 28.$ \\
 DELPHI & $189. \pm 48.$ &  $160. \pm 25.$  &  $244. \pm 37.$ \\
  L3    & $148. \pm 63.$ &  $176. \pm 35.$  &  $233. \pm 49.$ \\
  OPAL  & $70.  \pm 51.$ &  $156. \pm 25.$  &  $143. \pm 30.$ \\
\hline
LEP  &  $153. \pm 25.$   &  $164. \pm 13.$ &  $183. \pm 17.$ \\
\hline
\end{tabular}
\end{center}
\end{table}

\begin{figure}
\psfig{file=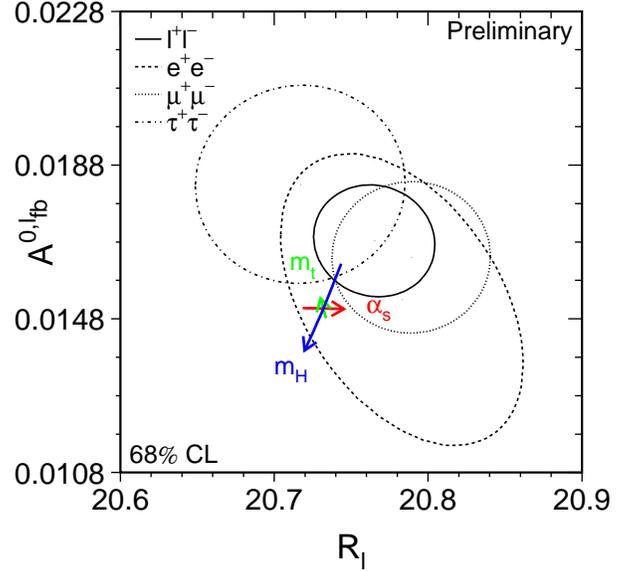,height=9.cm} 
\caption{$68\%$ probability contours in the $(\Rl,\Afbzl)$ plane with and
without the assumption of lepton universality. The SM prediction shown by the 
arrows was computed for $\Mt=173.8 \pm 5.0~GeV$, $90~GeV < \MH<1000~GeV$ and 
$\alfmz = 0.119 \pm 0.002$.}
\label{fig:afbl}
\end{figure}

Figure~\ref{fig:afbl} shows for each lepton species and for the combination 
assuming lepton universality, the resulting $68\%$ probability contours in the
 $(\Rl,\Afbzl)$ plane. 
\section{Derived quantities}
\begin{figure}
\psfig{file=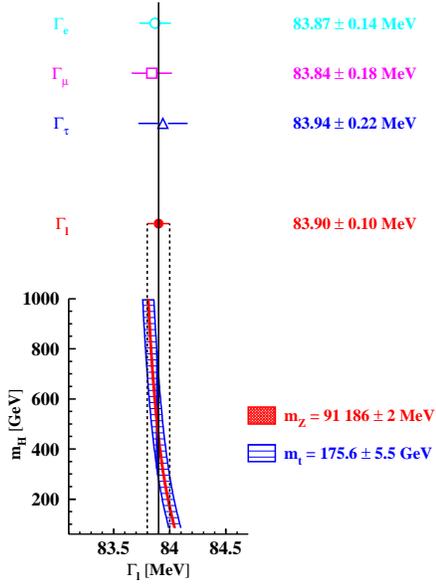,height=9.cm} 
\caption{Partial leptonic width values from combined LEP experiments.}
\label{fig:guni}
\end{figure}

The partial decay widths into hadrons and leptons can be obtained
by parameter transformation from the original nine and five parameters.
Since the total Z width is the sum of all partial widths, the decay width
into invisible particles,
$\Ginv = \GZ  - \Ghad - \Gll\;( 3+\delta_m )$,
can also be determined; here, $\delta_m = -0.0023$ is a small correction
which accounts for the $\tau$ mass effect. The results are summarised in 
Figure~\ref{fig:guni}. The comparison of the partial decay widths of the Z
into $e$, $\mu$ and $\tau$ shows good consistency with lepton universality.
To check whether the invisible width is completely explained by decays into
the three neutrino species, the ``number of neutrinos'', $N_\nu$
can be calculated according to
\begin{equation}
\frac{\Ginv}{\Gll} =
 N_\nu \left(\frac{\Gamma_\nu}{\Gll}\right)_{SM}\;.
\end{equation}
The Standard Model value for the ratio of the partial widths to neutrinos
and to charged leptons is $1.991 \pm 0.001$, where the uncertainty arises
from variations of the top quark mass within $\Mt=174 \pm 5~GeV$ and of 
the Higgs mass within $70~GeV < \MH<1000~GeV$.
The number of neutrinos from all experiments are shown in Table~\ref{tab:nnu}.
\begin{table}
\begin{center}
\caption{Number of light neutrino species detemined by each experiment and combined result.}
\label{tab:nnu}
\vspace{0.1cm}
\begin{tabular}{|c |c|}
\hline
 ALEPH  &  $2.993 \pm 0.015$ \\
DELPHI &   $2.988 \pm 0.018$ \\
L3 &  $3.005 \pm 0.018$ \\
OPAL & $2.988 \pm 0.015$ \\
\hline
 LEP  & $2.994 \pm 0.011$ \\
\hline
\end{tabular}
\end{center}
\end{table}    
\section{Conclusion}
The four LEP experiments have provided very precise measurements of the 
lineshape parameters and Forward-Backward asymmetries. Many results are still
preliminary but will become final soon. They will provide, together with 
results from SLD and NUTEV experiments further constraints on the Standard 
Model.
\section*{Acknowledgements}
 The averaging of the results was done inside the four-LEP experiments 
electroweak group. The impressive achievement of the LEP energy calibration 
group is also to be aknowledged. Therefore it is a pleasure to thank all my
colleagues from the two groups who helped me in preparing this talk.

\end{document}